\documentclass[proceedings, preprint]{rmaa}

\usepackage{paralist}

\usepackage{psfrag,color}

\SetYear{2013}
\SetConfTitle{Magnetic Fields in the Universe IV}

\title{Hall Equilibria: Solutions with toroidal and poloidal magnetic fields in Neutron Star Crusts} 

\author{
  K.N.~Gourgouliatos,\altaffilmark{1,2} 
  A.~Cumming,\altaffilmark{1}
  M.~Lyutikov\altaffilmark{3,4}
  and A.~Reisenegger\altaffilmark{5}}

\altaffiltext{1}{Department of Physics, McGill University, 3600 rue University, Montr\`eal, Qu\'ebec H3A 2T8, Canada (kostasg@physics.mcgill.ca).}

\altaffiltext{2}{Centre for Research in Astronomy in Quebec Fellow.}

\altaffiltext{3}{Department of Physics, Purdue University, 525 Northwestern Avenue, West Lafayette, IN 47907-2036, USA.}

\altaffiltext{4}{The Canadian Institute for Theoretical Astrophysics, University of Toronto,  60 St. George Street Toronto, Ontario, M5S 3H8 Canada.}

\altaffiltext{5}{Departamento de Astronom\'ia y Astrof\'isica, Facultad de F\'isica, Pontificia Universidad Cat\'olica de Chile, Av. Vicu\~na Mackenna 4860, 782-0436 Macul, Santiago, Chile.}

\shortauthor{K.N.~Gourgouliatos et al.}
\shorttitle{Hall Equilibria}

\listofauthors{K.N.~Gourgouliatos, A.~Cumming, M.~Lyutikov, \& A.~Reisenegger}
\indexauthor{Gourgouliatos, K. N.}
\indexauthor{Cumming, A.}
\indexauthor{Lyutikov, M.}
\indexauthor{Reisenegger, A.}

\abstract{We present Hall equilibrium solutions for neutron stars crusts containing toroidal and poloidal magnetic field.  Some simple cases are solved analytically while more complicated configurations are found numerically through a Gauss-Seidel elliptic partial differential equation solver. }

\resumen{}

\addkeyword{Neutron Stars}
\addkeyword{MHD}
\addkeyword{Methods: Analyical}
\addkeyword{Methods: Numerical}

\begin{document}
\maketitle

\section{Neutron Star Crust}

The crust is the upper layer of a neutron star ($\sim$1km) and consists of a highly conducting crystal lattice. Lorentz forces exerted by the magnetic field are balanced by the elastic crust and the field evolves through Hall effect \citep{Goldreich:1992}.

\section{Hall Evolution}

In the Hall effect, the magnetic field is advected by the electric current, which is carried by electrons. Ohmic diffusion, for neutron star crust applications, is much slower and is treated as a secondary effect.
\begin{eqnarray}
\frac{\partial \bf{B}}{\partial t} = -\underbrace{\frac{c}{4 \pi {\rm e}}\nabla \times \left( \frac{\nabla \times \bf{B}}{n_{\rm e}}\times \bf{B}\right)}_\textrm{Hall term} \nonumber \\
- \underbrace{\frac{c^{2}}{4\pi} \nabla \times \left(\frac{1}{\sigma} \nabla \times \bf{B}\right)}_\textrm{Ohmic term}
\end{eqnarray}
In a system dominated by Hall drift, equilibrium occurs when the Hall term is equal to zero. The field might settle into this state after several Hall timescales. We solve for Hall equilibrium inside the neutron star crust while requiring that the field connects to an external vacuum dipole field.

\section{Analytical Solutions}

\begin{figure}
\centering
\includegraphics[width=0.46\columnwidth]{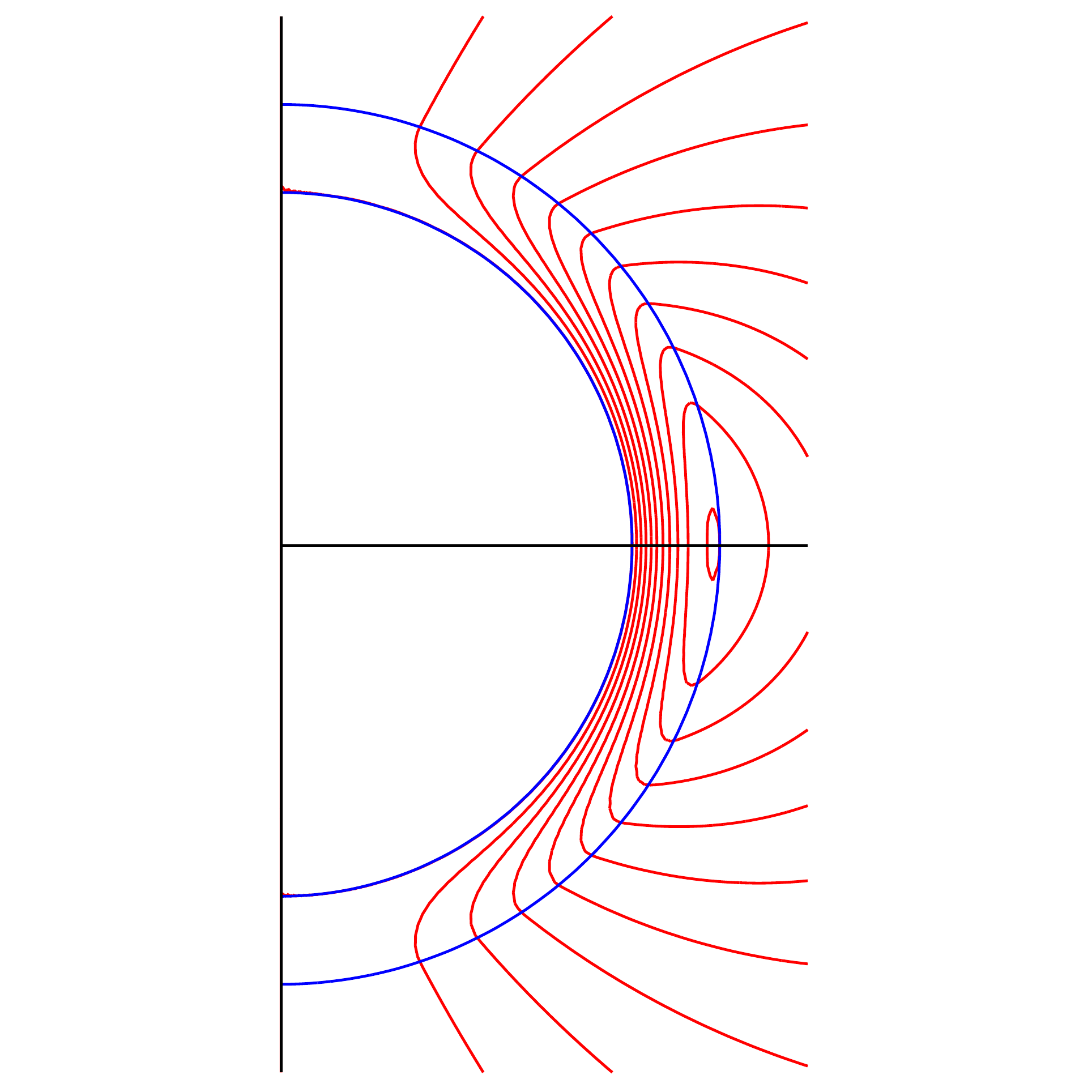}
 \includegraphics[width=0.46\columnwidth]{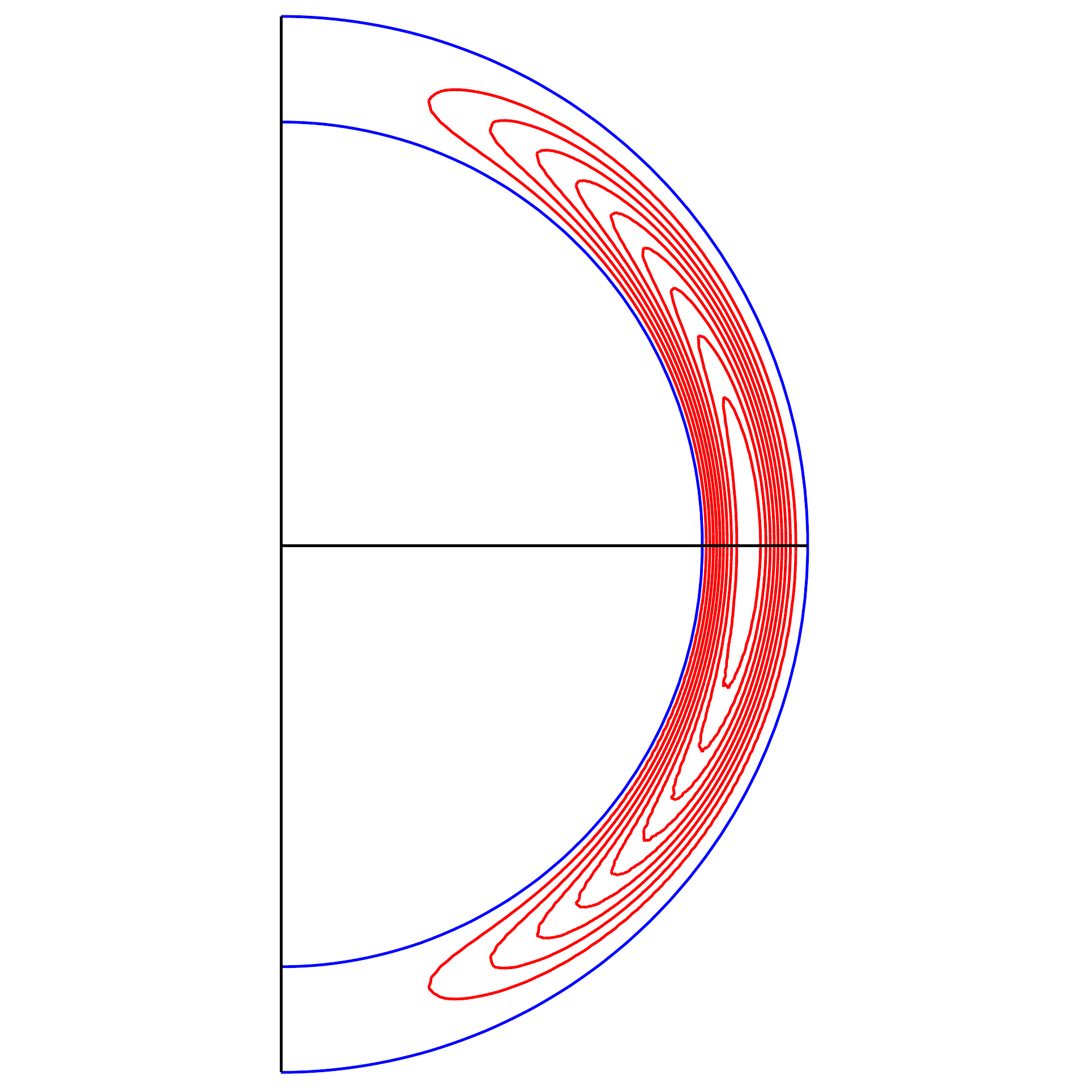}
\caption{Meridional sections of the star showing surfaces of constant $\Psi$ which are also the poloidal field lines. Left: The purely poloidal field solution corresponding to $S \propto \Psi$ and $I=0$. Right: The fully confined mixed poloidal and toroidal field corresponding $S\propto \Psi$ and $I\propto \Psi$. In both plots $n_{e}=$const.~ and the thickness of the crust is 0.2, assuming a stellar radius of unity.}
\label{Fig:1}
\end{figure}

We write the magnetic field in terms of the poloidal flux function $\Psi$ and the poloidal current $c I/2$:
\begin{eqnarray}
{\bf B} =\nabla {\Psi} \times \nabla \phi + I\nabla \phi\,.
\end{eqnarray}
Neglecting the Ohmic term and demanding that $\partial_{t} {\bf B}=0$, we obtain the Grad-Shafranov equation. Defining the Grad-Shafranov operator $\Delta^{*}=\partial^{2}_r +\frac{\sin^{2}\theta}{r^{2}}\partial_{\theta}\left(\frac{\partial_{\theta}}{\sin\theta}\right)$, Hall equilibria are given by
\begin{eqnarray}
\Delta^{*}\Psi + II'+r^{2}\sin^{2}\theta n_{\rm e} S'=0\,, 
\end{eqnarray}
where a prime denotes differentiation with respect to $\Psi$. $I$ and $S$ are arbitrary functions of $\Psi$. Analytical solutions which connect smoothly to a vacuum field outside the star are possible for linear choices of $S\propto \Psi$ and $I=0$. These fields have the angular structure of a dipole and correspond to uniform rotation of the electron fluid  \citep{Cumming:2004}. Analytical solutions for fields with mixed poloidal and toroidal components fully confined within the star are possible for $I\propto \Psi$ and bear similarities with previously known MHD solutions \citep{Gourgouliatos:2010}, see figure \ref{Fig:1}.

\section{Numerical Solutions}

\begin{figure}
\centering
\includegraphics[width=\columnwidth]{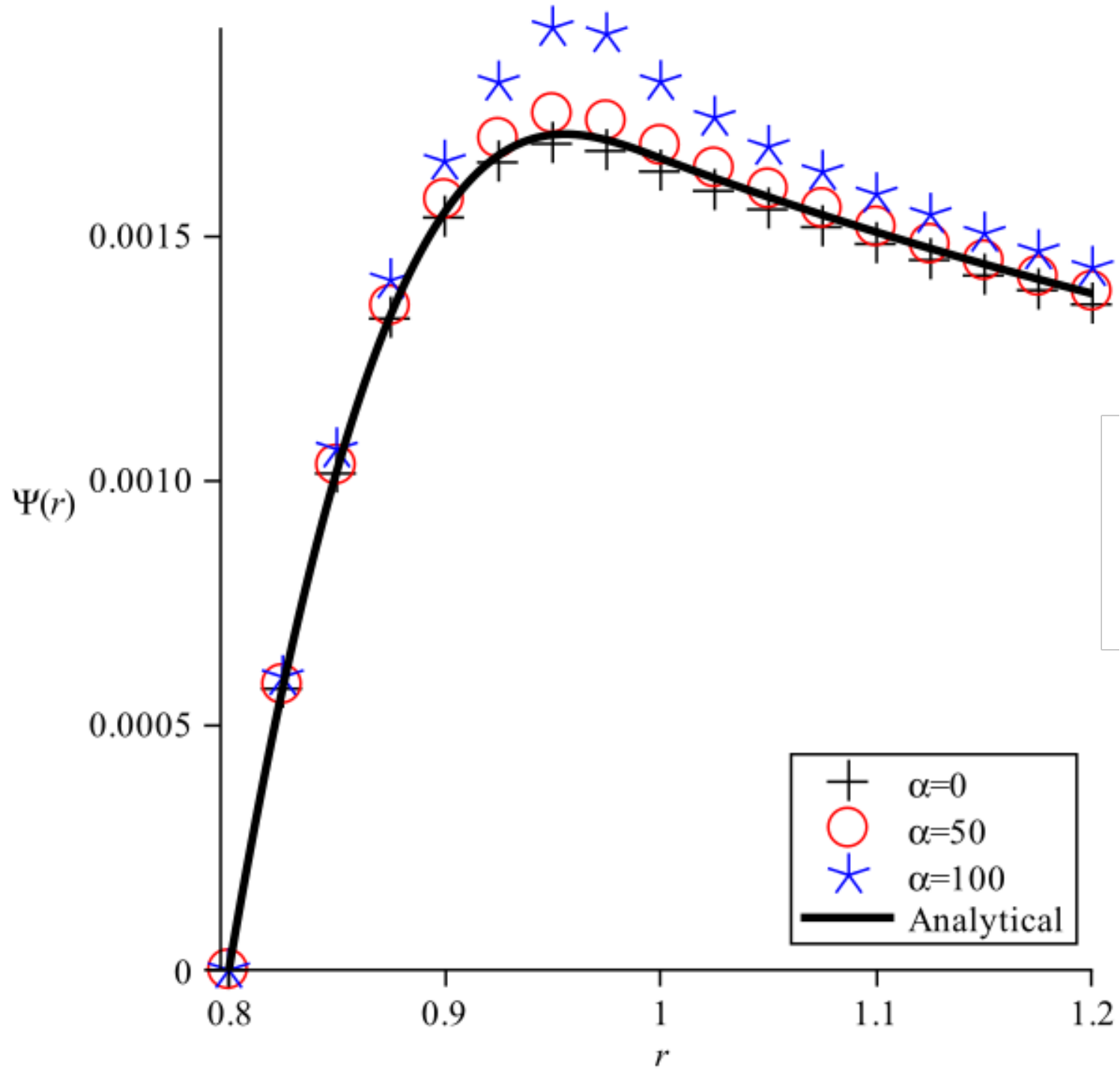}
\caption{Plot of the magnetic flux function at the equator $\Psi(r, \pi/2)$ as a function of the radius of the star found numerically. $I$ is chosen to be $I=\alpha(\Psi-\Psi_{0})^{1.1}$ where $\Psi_{0}$ is the value at the equator, $S'=$const., the thickness of the crust is $0.2$ and $n_{e}=1-r^{2}$. The solid line is the analytical solution, the black crosses is the numerical solution for the same case ($\alpha=0$) while the red circles and the blue stars correspond to $\alpha=50$ and $\alpha=100$ respectively.}
\label{Fig:2}
\end{figure}

\begin{figure}
\centering
\includegraphics[width=0.46\columnwidth]{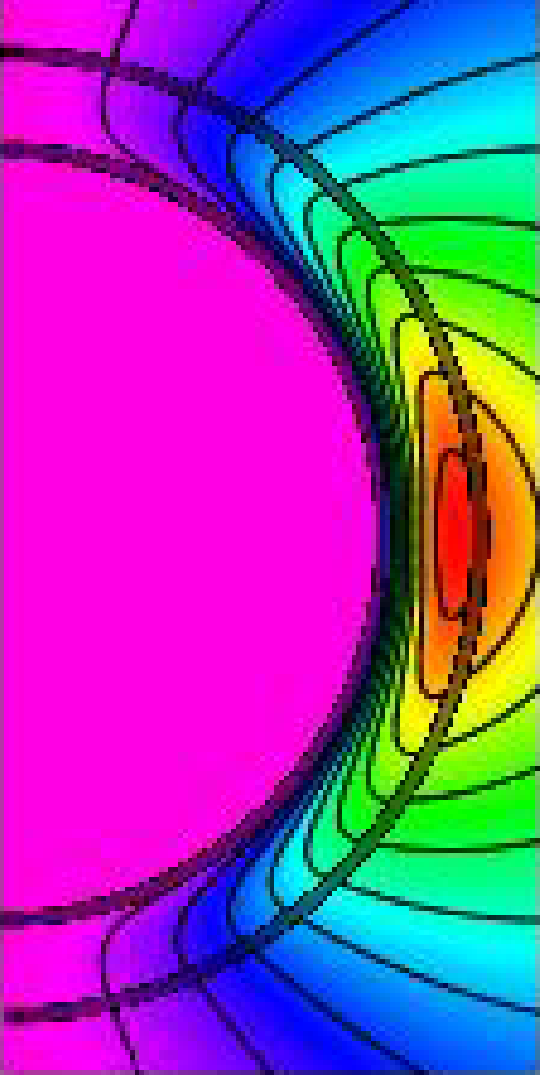}
\includegraphics[width=0.46\columnwidth]{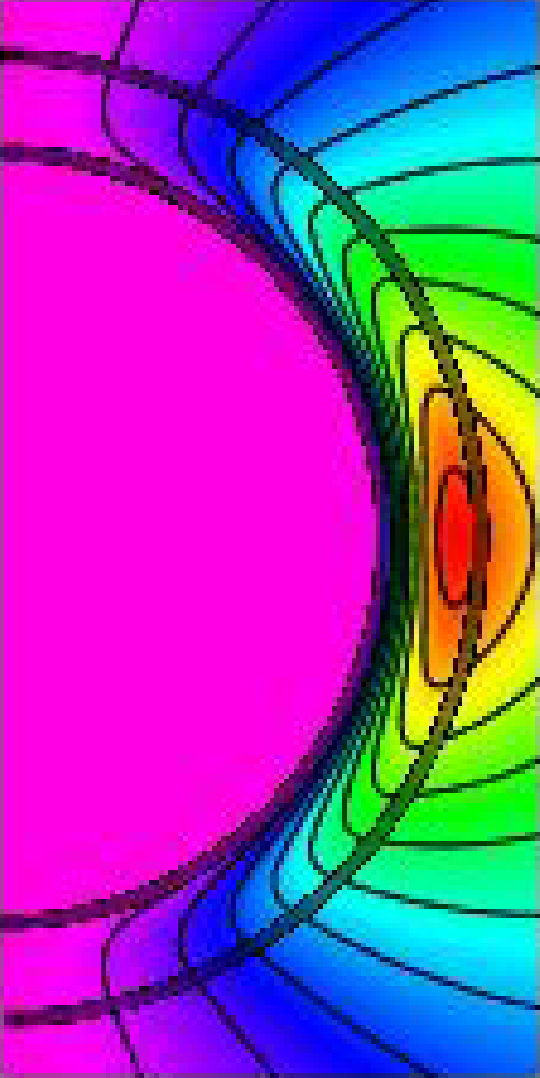}
\caption{Plots of the poloidal field lines with toroidal field. The colour represents the value of $\Psi$. The left plot corresponds to $\alpha=15$ and the right to $\alpha=45$. The toroidal field is hosted in the enclosed red loops. The choices of parameters are as in figure \ref{Fig:2}.}
\label{Fig:3}
\end{figure}

To avoid the drastic simplifications of analytical solutions, we numerically solve the Grad-Shafranov equation. The azimuthal field is confined to closed tori as the external vacuum cannot support currents. The numerical scheme reproduces the analytical results and has some similarities with the poloidal fields found in Hall simulations \citep{Pons:2007, Kojima:2012}. The solutions are plotted in figures \ref{Fig:2} and \ref{Fig:3}.

\section{Neutron Star Magnetic Evolution}

The solutions presented, subject to their stability, may represent long-term states in the magnetic evolution of neutron stars. Despite their similarities, they do not coincide with MHD equilibria, thus a phase transition during the formation of a neutron star from a fluid to a solid state will be followed by Hall evolution that may lead to Hall equilibrium. The Hall equilibrium will be slowly decaying by Ohmic diffusion from the start. Eventually, when this effect has already caused a substantial reduction of the field strength, the Ohmic term will dominate over the Hall term, and then the field structure will be essentially that of the lowest-order Ohmic mode rather than the Hall equilibrium. A more extended and in-depth discussion of this work will be published soon (Gourgouliatos et al., in preparation).


\begin{thebibliography}

\bibitem[Cumming et al.(2004)]{Cumming:2004} Cumming, A., Arras, P., \& Zweibel, E.\ 2004, \apj, 609, 999 
\bibitem[Gourgouliatos et al.(2010)]{Gourgouliatos:2010} Gourgouliatos, K.~N., Braithwaite, J., \& Lyutikov, M.\ 2010, \mnras, 409, 1660 
\bibitem[Goldreich \& Reisenegger(1992)]{Goldreich:1992} Goldreich, P., \& Reisenegger, A.\ 1992, \apj, 395, 250 
\bibitem[Kojima \& Kisaka(2012)]{Kojima:2012} Kojima, Y., \& Kisaka, S.\ 2012, \mnras, 421, 2722 
\bibitem[Pons \& Geppert(2007)]{Pons:2007} Pons, J.~A., \& Geppert, U.\ 2007, \aap, 470, 303 
\bibitem[Shafranov(1966)]{Shafranov:1966} Shafranov, V.~D.\ 1966, Reviews of Plasma Physics, 2, 103 
\end{thebibliography}
\end{document}